\documentclass{article}

\usepackage{arxiv}
\usepackage{graphicx}

\usepackage{algorithm}
\usepackage[noend]{algpseudocode}

\usepackage{float}

\usepackage[utf8]{inputenc} % allow utf-8 input
\usepackage[T1]{fontenc}    % use 8-bit T1 fonts
\usepackage{hyperref}       % hyperlinks
\usepackage{url}            % simple URL typesetting
\usepackage{booktabs}       % professional-quality tables
\usepackage{amsfonts}       % blackboard math symbols
\usepackage{nicefrac}       % compact symbols for 1/2, etc.
\usepackage{microtype}      % microtypography
\usepackage{lipsum}

\title{NFTracer: A Non-Fungible Token Tracking Proof-of-Concept Using Hyperledger Fabric}

\author{
  Mustafa Bal \\
  Department of Computer Science\\
  Harvard University \\
  Cambridge, MA 02138 \\
  \texttt{mbal@college.harvard.edu} \\
   \And
 Caitlin Ner \\
  Department of Computer Science\\
  Harvard University \\
  Cambridge, MA 02138 \\
  \texttt{caitlinner@college.harvard.edu} \\
}

\begin{document}
\maketitle

\begin{abstract}

Various start-up developers and academic researchers have investigated the usage of blockchain as a data storage medium due to the advantages offered by its tamper-proof and decentralized nature. However, there have not been many attempts to provide a standard platform for virtually storing the states of unique tangible entities and their subsequent modifications. In this paper, we propose NFTracer, a non-fungible token tracking proof-of-concept based on Hyperledger Composer and Hyperledger Fabric Blockchain. To achieve the capabilities of our platform, we use NFTracer to build an artwork auction and a real estate auction, which vary in technical complexity and demonstrate the advantages of being able to track entities and their resulting modifications in a decentralized manner. We also present its accompanying modular architecture and system components, and discuss possible future works on NFTracer.

\end{abstract}

% keywords can be removed

\section{Introduction}

It is a major concern to track the history of entities that carry and hold fluctuating value over time. Whether this may be governments guaranteeing title deeds or private companies hosting online listings of tangible entities, Central Information Providers (CIPs) exist to act as a source of information and as a place of exchange. CIPs provide an implicit promise that the information they hold is secure and accurate, which all participants in contact with the CIPs must agree by before taking part in the system. However, CIPs are by nature centralized, meaning there is only one point of failure for the whole system to break down. This point of failure can be easily breached by various organizations; this may be a cyber-attack by a malicious group of hackers aiming crippling the network that is hosting the digital information or a domestic government institution deciding that a given CIP is no longer in their interest. 

Even past the destroying of data argument, people with malicious intentions or those who are unaware they are entering wrong data can insert invalid and inaccurate information into the CIPs data servers. And because of the single point of failure being the data servers themselves, there would be no easy way of fact-checking the entered data, thereby resulting in the declaration of faulty information as valid and genuine.  As such, there is an inherently serious problem of guaranteeing the security and validity of information about real-world entities online. In direct response to this problem, we introduce NFTracer, our decentralized blockchain solution to store and track the data of entities in a real-world context.

The main objective of this paper is to provide a blockchain based platform for non-fungible token tracking. We contribute to the literature of Blockchain in three ways: 

\begin{enumerate}
    \item We bring together various literature to explain the possibility of achieving true decentralized non-fungible token storage 
    \item We develop a comprehensive system using the Hyperledger Fabric's blockchain network architecture to execute decentralized non-fungible token tracking 
    \item We analyze and demonstrate how the Fabric can provide solutions to the art and real estate auction systems
\end{enumerate}

In this paper, we first explore relevant background topics such as the difference between permissioned and permissionless blockchain, the Hyperledger Fabric transaction flow, the difference between Hyperledger’s own Chaincode as compared to pre-existing smart contracts like that of Ethereum, and details on non-fungible tokens. We then transition to related works that have influenced the researching, designing and manufacturing of NFTracer including its purpose and its back-end implementation. Next, we review NFTracer’s system implementation, going into detail on its architecture and inner framework components. We give two possible applications of the usage of NFTracer that vary in size and complexity, where the art auction model is the smaller and less technically complex version and the real estate model is the larger and more technically complex. Lastly, we deliver our conclusions on NFTracer and discuss possible future works that may result directly from this paper.

\section{Background}
\label{sec:headings}

\subsection{Permissioned and Permissionless Blockchain}

Blockchain technology originally emerged from the development of Bitcoin as a distributed, immutable ledger that is maintained and verified among a network of peers \cite{nakamoto2008bitcoin}. Several industries have since then explored the underlying peer to peer technology and its ability to create cost-efficient and decentralized business networks \cite{androulaki2018hyperledger}.

In permissionless or public blockchains, any agent can participate and send transactions while maintaining an anonymous identity. Permissionless blockchains often include an embedded, native cryptocurrency and “proof of work” (PoW) consensus mechanisms. Permissioned blockchains such as Hyperledger Fabric require an additional layer of authentication to run among a set of identified participants.  This offers security to a group of entities that does not completely trust one another but wants to achieve a common objective such as exchanging information or goods. A permissioned blockchain can use Byzantine-fault tolerant (BFT) consensus among all members of a network.

The Fabric includes a membership identity service that administers user IDs and verifies participants on the network \cite{Fabric2019}. Members therefore know each other’s identity but still maintain privacy and confidentiality because they do not know what the other members are doing. 

\begin{figure}
  \centering
  \includegraphics[scale=0.6]{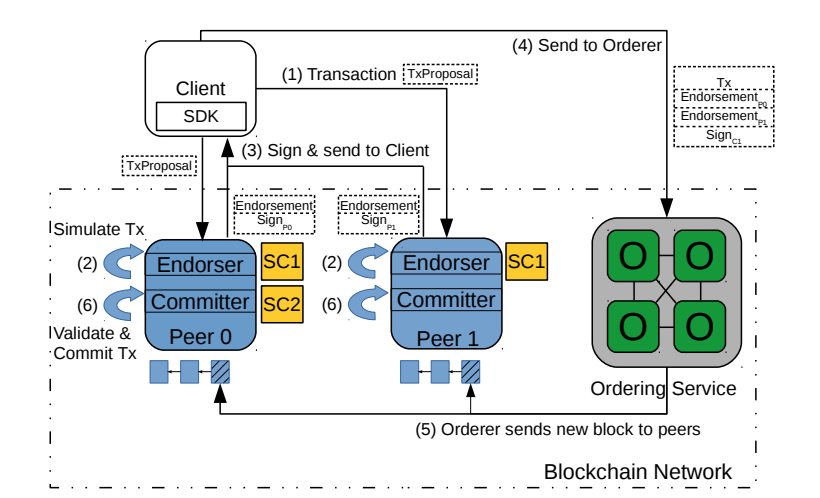}
  \caption{Hyperledger Transaction Flow
  \cite{sukhwani2018}}
  \label{fig:hyperledger}
\end{figure}

\subsection{Hyperledger Fabric}
The Hyperledger Project was founded by the Linux Foundation in 2015 to offer a collaborative approach to develop blockchain technologies in enterprise contexts. Hyperledger Fabric is an open source software project licensed under Apache License, Version 2.0. 

The Hyperledger Fabric architecture differs from the existing architecture of several other smart-contract based blockchain platforms which resemble a traditional state-machine replication approach \cite{Schneider:1990:IFS:98163.98167}.  Ethereum and Quorum, for example, have an order-execute architecture that includes a consensus protocol to first order the transactions and then each peer executes the transactions. 

Hyperledger Fabric, on the other hand, separates the ordering and execution of transactions. Hyperledger Fabric is a modular, permissioned blockchain that supports the consistent execution of distributed applications and divides the transaction flow into three steps: (1) executing a transaction and checking its correctness (2) ordering through a BFT consensus protocol and (3) transaction validation\cite{Fabric2019}. The three steps can be performed on different peers which is a significant shift away from state-machine replication and improves scalability and modular consensus implementations. Hyperledger Fabric instead follows an execute-order-validate architecture which allows transactions to be executed prior to consensus. A detailed transaction flow can be found in Figure 1.

\subsubsection{Nodes}
Blockchain networks contain several nodes that communicate with one another to process transactions. Hyperledger Fabric is a permissioned network therefore all nodes on the network have an identity that is issued by the MSP membership service provider that is associated with an organization. The Fabric has three types of nodes: peers, orderers, and clients.

Peers execute the transactions and receive state updates through the form of blocks of transactions. After receiving the latest block, the peer validates the transactions and adds the changes onto a local version of the ledger and affixes the block onto the blockchain. Peers can also endorse transactions as endorsers by executing Hyperledger's smart contract or "chaincode" and attaching a cryptographic signature to be sent back to the client. Every organization can have multiple peers with one configured as an anchor peer that collects the blocks from the order and broadcasts them to other blocks through peer to peer gossip protocol \cite{Fabric2019}.  

The orderer's request for all the transactions in the network, propose new blocks, and seek consensus. Orderers also preserve the consortium or list of organizations that can create channels.

The clients are bodies that act specifically for an end-user. Clients send the transaction proposals that are received by peers and sends the endorsed transaction to the ordering service. The client is associated with the peer in its organization so that it can be aware of committed transactions.

\subsection{Hyperledger Chaincode versus Ethereum Smart Contracts}

Smart contracts are programmable contractual clauses that automate and define rules for inter-party transactions without trusted intermediaries \cite{szabo1996smart}. Ethereum, a widely used distributed blockchain platform to deploy decentralized applications, has key differences in its smart contracts compared to Hyperledger Fabric’s “chaincode” smart contracts. 

\begin{figure}
  \centering
  \includegraphics[scale=0.6]{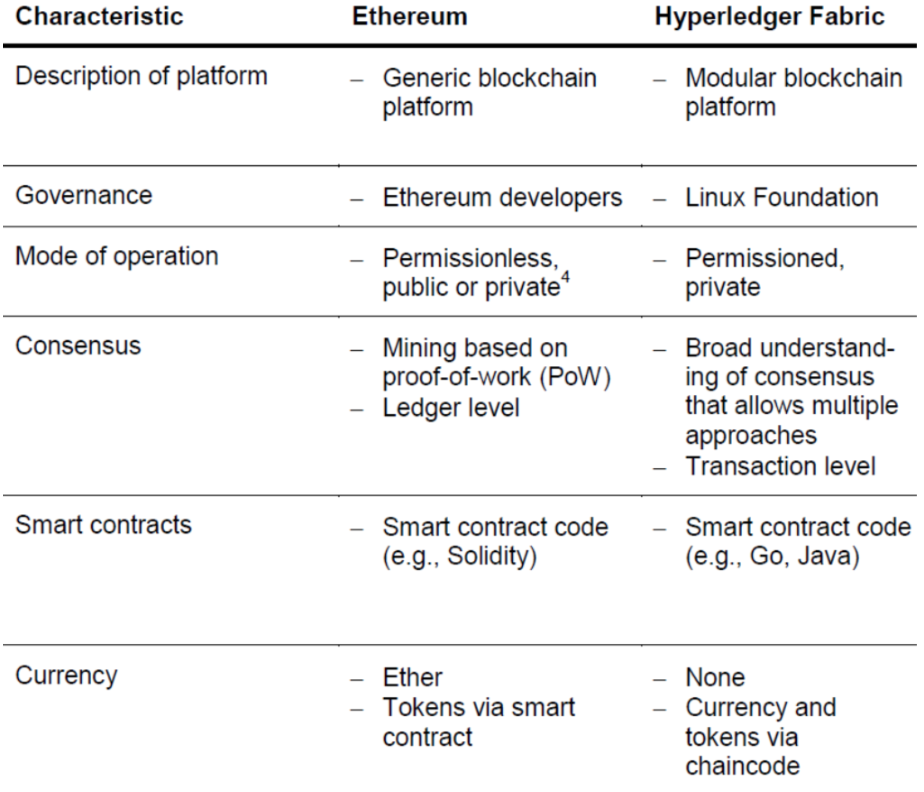}
  \caption{Major differences between Ethereum and Hyperledger Fabric \cite{valenta2017comparison}}
  \label{fig:ethvshyp}
\end{figure}

Fabric allows multiple chaincodes to be created on the blockchain with its own code and database. The chaincode’s database can only be altered by its code written by an untrusted developer rather than through transactions. Ethereum smart contracts, on the other hand, are executed by the Ethereum Virtual Machine (EVM) that are transparent and visible to all participants on the network. Ethereum uses the Solidity language to write smart contracts while Hyperledger Fabric utilizes mainstream programming such as Golang or Java. Hyperledger Fabric has a different approach compared to Ethereum in that if a “client” node wants to send a message to another chaincode, it has to send it to “endorser” nodes which execute the chaincodes independently, then determine the message’s effect on the database of the chaincode. An “endorsement” includes the endorser’s response and digital signature. A transaction can then be broadcasted if the client receives a sufficient number of endorsements \cite{FabricPeers}. 

\subsection{Non-Fungible Tokens}

Cryptographic assets can vary in terms of fungibility or non-fungibility. Fungibility is a characteristic of a token that dictates whether items or quantities of the similar type can be entirely interchangeable during exchange or utility \cite{understandingNFTs}. A non-fungible token (NFT) is a cryptographically unique, non-replicable token \cite{understandingNFTs}. This is in contrast to fungible tokens which can be replaced by other things in the real world such as currency or stocks which can be split and exchanged. A limitation to fungible tokens are that many valuable things, such as art work or real estate, cannot be divided and replaced. Non-fungible tokens therefore offer an improved system to record and keep track of the ownership of unique assets. The ownership along with any relevant information can be recorded and permanently kept on the blockchain.

Fungible tokens differ from non-fungible tokens in term of interchangeability, uniformity, and divisibility. Fungible tokens can be exchanged for other items including one currency for another of the same currency. They are also identical to one another and can be divided into smaller units and not affect its value. A 100 dollar bill divided into 100 one dollar bills would be equal and therefore fungible. 

Non-fungible tokens as proposed in this paper cannot be replaced with another non-fungible token of a similar type e.g. a non-fungible token of one fine artwork cannot be exchanged with a non-fungible token of another fine artwork. A non-fungible token cannot be divided e.g. you cannot divide a constructed real estate property into different parts. Each non-fungible token is unique with distinctive information and attributes that make them impossible to interchange. Each non-fungible asset is unique and differs from others. Certain artwork may have the same dimensions or artistic style, but each one has a unique artist, date of creation, and gallery identifier and therefore does not make it easily transferable. 

Due to the non-fungible nature of assets such as fine art and real estate, we structure our system accordingly. Non-fungible tokens in relation to these objects then become unique investments because they are tied to a physical object and ownership of the assets cannot be faked once put on the blockchain.

\section{Related Works}

ERC-721 tokens \cite{entriken2018erc} were developed in 2017 and allowed developers to tokenize the ownership of any arbitrary data on the Ethereum blockchain. Each ERC-721 non-fungible token is tied to a distinct identifier that is unique to each owner. 

CryptoKitties \cite{cryptokitties} popularized the ERC-721 standard. CryptoKitties are digital collectibles that are indivisible and unique. Each CryptoKitty was represented in the form of an ERC-721 token stored in an Ethereum smart contract with specific phenotypes determined by genotypes. The ability to track ownership in a decentralized blockchain made it desirable in terms of rare collectibles and reveals opportunities for future non-fungible tokens to be applied to more valuable assets. 

Blass and Kerschbaum \cite{blass2018strain} introduced Strain, a protocol that executed sealed-bid auctions on top of blockchains to offer confidentiality against fully malicious parties. The authors proposed a two party comparison mechanism that was executed between bidders. All bidders are able to verify the mechanism through zero knowledge proofs.

\section{Application Areas}

In this section, we briefly explore two application areas that will be modeled by our versatile PoC: Art and Real Estate Auction. We go into the reasons why a decentralized action market is necessary, and argue how NFTracer will alleviate our concerns on their current states.

\subsection{Application 1: Art Auctions}
The fine art market is opaque as the demand for masterpieces has significantly increased while the supply stays limited. The lack of a transparent database of the world’s greatest artworks has led to forgeries in both the actual artwork and the supporting documentation. For example, The Fine Arts Expert Institute (FAEI) in Geneva revealed that over 50 percent of their artworks were either listed as the wrong artist or forged \cite{artnetnews_2014}. There is a need for a better system to authenticate art pieces, establish ownership, and transfer rights. 

The current technique of transferring art is highly opaque which makes it difficult to track the movement of item as well as provenance, which assigns the artwork to a known artist with a detailed history. The transaction records within the art industry are often still paper-based which leads to a multitude of risks. A decentralized system can keep a list of records that are secured through cryptography and upheld through its fragmented nature. We apply the NFTracer system specifically with art auctions in order to track and verify authenticity according to the cryptographic signatures and timestamps during the transfer of ownership of the artwork.  

\subsection{Application 2: Real Estate Auctions}
Real estate property is an illiquid asset class that also faces a lack of transparency and several hidden costs and regulations. Real estate has expensive transaction costs, land use regulations, long lasting renovations, and slow responses due to supply changes \cite{dijkstra2017blockchain}. Real estate assets have the primary characteristics of immobility and heterogeneity. The overarching market for buying and selling real estate tends to be illiquid, localized and segmented with privately negotiated transactions along with expensive transaction costs due to the participation of a considerable amount of trusted third parties.

A decentralized action market can benefit the real estate industry in reaction to a greater need for transparency by making more of the information digitized and available. The lifecycle of the real estate property can be transferred onto the blockchain to create a system where each property has its own digital information about the specific asset including title registration, recent owners, and sales prices. NFTracer can improve transparency by offering an immutable record of ownership and performance which can allow regulators and future clients to better understand any risks affiliated with the real estate and future potential for investment. 

\begin{figure} 
  \centering
  \includegraphics[scale=0.7]{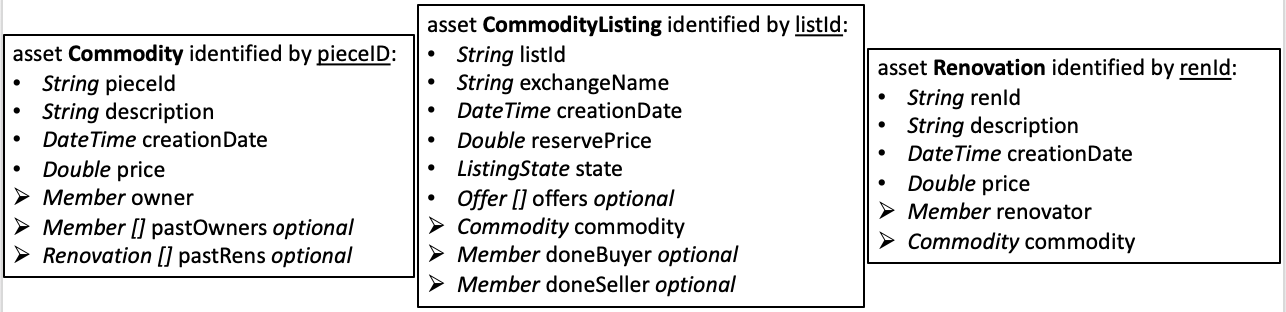}
  \includegraphics[scale=0.7]{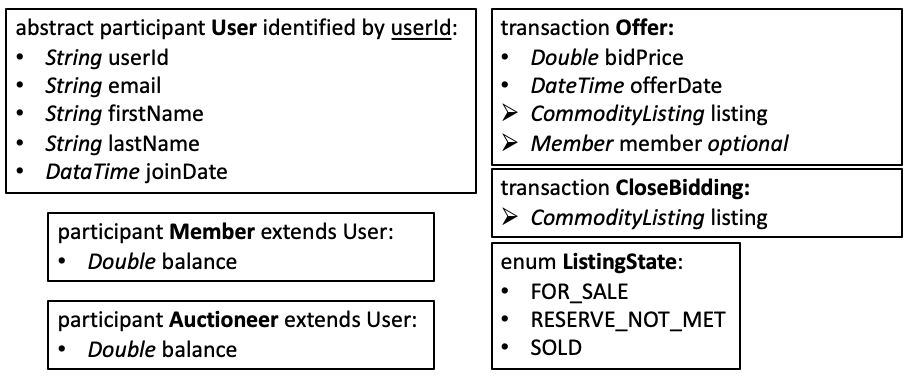}
  \caption{The classes of Assets, Participants, Transactions, and Enum in NFTracer. Features of classes preceded a bullet are direct data fields of the class, and features of classes preceded by an arrow are data fields that also function as foreign keys to other class types.}
  \label{fig:figAssets}
\end{figure}

\section{Model Overview}

In this section, we briefly explain the essential modules in the design of NFTracer. Our Hyperledger Fabric platform consists of three main entities: Assets, Participants, and Transactions. We also have the Enum entity to provide a clear representation of the state of the auction listing. We give the implementation specifics of these classes in Figure \ref{fig:figAssets}.

As with NFTracer we are tracking real-world tangible items in an auction environment, we have made three Asset types: Commodity, CommodityListing and Renovation. The Commodity Asset type will represent the real-world tangible items, and will include information such as its Id, description, ideal price and its owner. The auctioneer will also be able to track the past owners of the Commodity and/or the renovations/modifications it may have undergone. This allows the auction environment to be tailored to the needs of the situation. For example, normal art auction do not track modifications done to the art piece as that would diminish the value of the piece and therefore will not make sense to track it. However, in a real estate auction scenario, it does make sense to track renovations done to the house, and would therefore be enabled. This Renovation Asset type can track the date of the renovation, the price it cost to make the renovation, the owner who renovated, etc... The CommodityListing Asset type allows us to list the Commodity Assets on an auction, and comes with its own attributes such as its list ID, the name of the exchange it is listed in, its reserve price, the offers that have already been made, and more.

With an abstract Participant Class type, we are able to represent members of NFTracer who may bid on and sell items in an auction, as well as auctioneers who are able to direct auctions by calling the time in which an auction session is over. With the Offer Transaction, a Member Participant may bid on an auction with their chosen price on their chosen CommodityListing. Lastly, the auctioneer is the only person who is allowed to call the CloseBidding Transaction that stops a given auction, where the item is sold if the highest bidder has bid more than the minimum reserve price set by the seller, or is not sold if the minimum reserve price hasn't been met.

\section{System Implementation}

We will now dive into the steps that each Participant will go through on NFTracer. We will focus on Buyers, Sellers and Auctioneers at the same time. We have made our repository available for viewing on Github \footnote{NFTracer Github Repo: https://github.com/mstfbl/nftracer}. The specific algorithms can be found in Chapter \ref{ch:alg}. 

\subsection{Phase 1: Buyers, Sellers and Auctioneers are inducted into NFTracer}

As shown in Algorithm \ref{alg:p1}, in the first phase NFTracer will first ready the system and auction environment. This will be done by including the buyers, sellers and auctioneer(s)  

\subsection{Phase 2: Sellers add their entities for sale}
As shown in Algorithm \ref{alg:p2}, the sellers must add the information of the commodity that they will sell.  This will be done by including the exchange name, commodity, and seller ID. 

\subsection{Phase 3: Buyers make their bids}
As shown in Algorithm \ref{alg:p3}, the buyer will bid on a commodity by specifying the CommodityListing and their bid price.

\subsection{Phase 4: Auctioneer ends the auction}
As shown in Algorithm \ref{alg:p4}, the auctioneer closes the bidding by choosing the highest bid and according to if the CommodityListing reserve price has been met.

\subsection{Phase 5 and Beyond: Assets are transferred, if auction is successful}
As shown in Algorithm \ref{alg:p5}, if the CommodityListing has the ListingState "SOLD", and if the "doneBuyer" attribute of the CommodityListing matches the proposed new owner, the Commodity is transferred to its new owner. Else, no change occurs. This cycle repeats for each new auction made. 
\section{System Analysis}

After implementing our NFTracer system, we set up multiple test runs to test for any bugs that might exist in our logic, check for any time running issues, and overall to make sure our implementation works correctly. We implemented separate test cases for both of our existing application implementation. In addition, we put extra care into making sure that only the users who are supposed to be able to access certain parts of the program can access those given parts, and that no malicious intentions would be immediately glaring to a possible hacker. For example, this involves ensuring that any member of the platform who is not the auctioneer would be able to call the end of bidding at a time that benefits the opportunistic bidder. Another case may be making sure the seller of a given entity does not have total control over when the bidding is declared complete, so that if there are no buyers willing to pay the minimum cost for the entity to be sold, the seller is not able to clog the system and leave their old item on the platform for too long.

\section{Algorithms}
\label{ch:alg}

\begin{algorithm} [H]
\caption{Creating an NFTracer auction platform}\label{alg:p1}
\begin{algorithmic}[1]
\Procedure{InitiateAuctionEnvironment}{$buyersLst,sellersLst,auctioneer $}
\State $Assert(buyersLst \neq [])$
\State $Assert(sellersLst \neq [])$
\State $Assert(auctioneer \neq NULL)$
\State $activeBuyers \gets buyersLst$
\State $activeSellers \gets sellersLst$
\State $activeAuctioneer \gets auctioneer$
\State $auctionId \gets startAuction(activeBuyers, activeSellers, activeAuctioneer)$
\State \textbf{return} $auctionId$
\State 
\EndProcedure
\end{algorithmic}
\end{algorithm}

\begin{algorithm} [H]
\caption{Seller entering info of Commodity to sell}\label{alg:p2}
\begin{algorithmic}[1]
\Procedure{createCommodityListing}{$exchangeName, myCommodity, sellerId$}
\State $ Assert(exists(exchangeName)) $
\State $ Assert(exists(myCommodity)) $
\State $listing \gets createCommodityListing(sellerId, exchangeName, myCommodity)$
\State \textbf{return} $submitToExchange(listing, exchangeName) $
\EndProcedure
\end{algorithmic}
\end{algorithm}

\begin{algorithm} [H]
\caption{Buyer makes a bid on a Commodity}\label{alg:p3}
\begin{algorithmic}[1]
\Procedure{makeBidOnCommodityListing}{$commodityListing, potentialBuyer, bidPrice$}
\State $ Assert(exists(commodityListing)) $
\State $ Assert(exists(potentialBuyer)) $
\State $ Assert(bidPrice > commodityListing.maxBid) $
\State $ sendBid(potentialBuyer, commodityListingm bidPrice)$
\State \textbf{return}
\EndProcedure
\end{algorithmic}
\end{algorithm}

\begin{algorithm} [H]
\caption{Auctioneer ends the auction}\label{alg:p4}
\begin{algorithmic}[1]
\Procedure{CloseBidding}{$commodityListing, Auctioneer$}
\State $ Assert(exists(commodityListing)) $
\State $ Assert(exists(Auctioneer)) $
\State $ AssertTrueAuctioneer(Auctioneer)) $
\If {$commodityListing.maxBid \geq commodityListing.reservePrice$}
    \State $ commodityListing.state \gets Enum.Sold $
    \State $ commodityListing.doneBuyer \gets commodityListing.maxBid.member $
\Else
    \State $ commodityListing.state \gets Enum.ReserveNotMet $
\EndIf
\State \textbf{return}
\EndProcedure
\end{algorithmic}
\end{algorithm}

\begin{algorithm} [H]
\caption{If auction is successful, assets are transferred}\label{alg:p5}
\begin{algorithmic}[1]
\Procedure{TransferAssetsIfAuctionSuccessful}{$commodityListing,highestBidder$}
\State $ Assert(exists(commodityListing)) $
\State $ Assert(exists(highestBidder)) $
\If {$commodityListing.state == Enum.Sold$}
    \State $ commodityListing.commodity.owner \gets highestBidder $
\EndIf
\State \textbf{return}
\EndProcedure
\end{algorithmic}
\end{algorithm}

\section{Conclusion and Next Steps}

In this paper, we have presented a non-fungible token tracking proof-of-concept by utilizing the Hyperledger Fabric, Hyperledger Composer and Hyperledger's Chaincode program. We have justified the existence of the problem of relying on a centralized entity for secure and reliable non-fungible token tracking. We then explained to the reader the necessary background information ranging from the difference between permissioned and permissionless blockchain to the pros and cons of both Hyperledger Chaincode and Ethereum Smart Contracts. Next we transitioned to go over works that are related to our NFTracer platform, and how implementation of a non-fungible token tracking proof-of-concept is different than the most famous usage of ERC0721 tokens so far, CryptoKitties. Then we talked in detail about our two intended implementations to represent the importance of tracking non-fungible entities and its potential it can bring to these auctions. We briefly went over important parts of Hyperledger's infrastructure that allow us to achieve a full decentralized and easily-scalable blockchain application, and how we implemented before walking through each step of the auction and non-fungible token tracking concept.    

For future work, we plan to explore more areas in which we can use NFTracer to track entities in the world with the non-fungible token logic. Ideal candidates for these areas are those that deal with "English auctions" (where members bid in higher and higher amounts until no other member can overbid the current-highest bidder) to facilitate the exchange of entities (such as car auctions). While our demonstrative platforms were implemented to work with English auctions, by implementing different profiles to realize the logic of auctions, we can also realize Dutch auctions (seller sets a high price, and lowers it down until a willing buyer appears) \cite{comment1991relative}, Vickrey auctions (all bids are submitted in secret and unaware of one another, the highest bidders wins the auction and pays the price offered by the second-highest bidder)\cite{ausubel2006lovely}, and more. 

\newpage

\bibliographystyle{unsrt}  
\bibliography{references}  %%% Remove comment to use the external .bib file (using bibtex).

\end{document}